\documentclass[pre,twocolumn,showpacs,endfloats]{revtex4}
\usepackage{graphicx,color,epsfig}
\begin{document}

\title{A pseudo-spectral method for the Kardar-Parisi-Zhang equation.}

\author{Lorenzo Giada}
\affiliation{International School for Advanced Studies (SISSA) and
  INFM Unit\'a di Trieste, v. Beirut 2-4, Trieste I-34014, Italy}
\email{giada@mpikg-golm.mpg.de}
\altaffiliation[Present address: ]{MPIKG Golm, D-14424 Potsdam, Germany}
\author{Achille Giacometti}
\affiliation{INFM Unit\'a di Venezia, Dipartimento di Chimica Fisica,
  Universit\'a di Venezia, Calle Larga Santa Marta DD 2137, I-30123
  Venezia-Italy}
\email[To whom correspondence should be sent: ]{achille@unive.it}
\author{Maurice Rossi}
\affiliation{Laboratoire de Mod\'elisation en M\'ecanique,
Universit\'e de Paris VI, 4 Place Jussieu, F-75252 Paris Cedex 05, France}

\date{\today}

\begin{abstract}
We discuss a numerical scheme to solve the continuum
Kardar-Parisi-Zhang equation in generic spatial dimensions.  It is
based on a momentum-space discretization of the continuum equation and
on a pseudo-spectral approximation of the non-linear term.  The method
is tested in $(1+1)$- and $(2+1)$- dimensions, where it is shown to
reproduce the current most reliable estimates of the critical
exponents based on Restricted Solid-on-Solid simulations.  In
particular it allows the computations of various correlation and
structure functions with high degree of numerical accuracy.  Some
deficiencies which are common to all previously used finite-difference
schemes are pointed out and the usefulness of the present approach in
this respect is discussed.
\end{abstract}
\maketitle
\section{Introduction}
\noindent 
Surface growth is a paradigmatic problem of nonequilibrium statistical
mechanics with widespread potential applications such as molecular
beam epitaxy, fluid flow in porous media or flame
fronts \cite{Barabasi95,Halpin-Healy95}.

For such extended systems, it is a rather intricate question how to
connect, in a direct way, microscopic interactions to the dynamics at
mesoscopic or coarse-grained scales.  Phenomenological models, based
on stochastic partial differential equations, selected according to
symmetry principles and conservation laws, are often capable to
reproduce various experimental data. The most well-known of these
models is the one introduced by Kardar, Parisi and Zhang (KPZ)
\cite{Kardar86}.  The equation introduced by these authors engendered
an enormous amount of work, that addressed the large number of issues
related to it \cite{Krug97}.  Yet, many fundamental properties of the
KPZ equation are still not well understood. Using renormalization
group (RG) theory, various authors attempted to estimate critical
exponents and the upper critical dimension
\cite{Frey94,Janssen97,Lassig95,Castellano98}.  The success of this
procedure has been limited, so far, by the difficulties of RG
techniques to reach the strong-coupling regime. An alternative route
to study stochastic partial differential equations, which yields an
easy and reliable access to critical exponents, hinges on the
so-called restricted solid-on-solid (RSOS) growth models
\cite{Kim89,Ala-Nissila94,Marinari00}.  This approach is based on
simple rules for deposition and diffusion of particles on a discrete
lattice, and it can be implemented in a very efficient and fast way on
a computer.  However, there is no {\it a priori} guarantee that a RSOS
model belongs to the universality class of the continuum equation,
albeit there is a general believe in this sense. Moreover, the
corresponding coupling parameters possess fixed values in the RSOS
case, a feature that prevents the exploration of the entire phase
diagram of the KPZ equation.

Given these premises, numerical integration appears as the most direct
and definite way to determine the universality class of a given
continuum equation.  So far, finite-differences methods have been
exploited in the context of the KPZ equation
\cite{Amar90,Moser91,Beccaria94}.  In this framework, Lam and Shin
\cite{Lam98} have shown the jeopardy of selecting an incorrect
discretization in the framework of finite-differences algorithms.  In
this work, we follow a different route and propose a numerical scheme
based on a pseudo-spectral representation.

Spectral methods, although almost routinely employed in fluid
mechanics \cite{Canuto88}, have not, so far, been used in the context
of stochastic equations except in Ref.~\cite{Giacometti01}. In this
latter work, two of us have introduced in the simple $(1+1)$ case, a
discretization algorithm based on a pseudo-spectral scheme which
outperforms classical finite differences on various respects. The
present work generalizes the spectral method to the 
non-trivial $(2+1)$ case, where no exact results are available and
where finite-differences methods may be
hampered by large discretization effects, as we will shortly discuss.
Extended numerical simulations for the $(2+1)$ case, are then
reported, along with comparisons with results based on
finite-differences schemes. The absence of uncontrolled discretization
effects also allow us to compute various correlation functions and
critical exponents in a rather precise and reliable way. Particularly
interesting is the computation of various height-height
correlation functions and of the structure factor, which were not
previously calculated.

The plan of the remaining of the paper is as follows. In section
\ref{sec:discretized}, we address the general features of
finite-differences discretizations and introduce the new
pseudo-spectral procedure.  Sect.~\ref{sec:relevant} contains
definitions of the various quantities employed in this work as well as
their pertinent scaling forms.  Results are presented in
Sect.~\ref{sec:case1} for the $(1+1)$-dimensional and in
Sect.~\ref{sec:case2} for the $(2+1)$-dimensional cases,
respectively. Conclusions and perspectives are included in
Sect.~\ref{sec:conclusions}.

\section{The discretized KPZ equation}
\label{sec:discretized}
\noindent
We consider a $d-$ dimensional substrate and denote with ${\bf x}$ the
$d-$ dimensional vector locating a point on it. The surface, which
grows on this substrate, is described at each time $t$ by its height
$h({\bf x},t)$.  In the continuum approximation, which disregards
overhangs, this dynamics generically satisfies a stochastic partial
differential equation
\begin{eqnarray}
\partial_t h  &=&  {\cal F}(h)+ \eta,
\label{eq1.1}
\end{eqnarray}
where ${\cal F}(h) $ is a functional containing various derivatives of
$ h({\bf x},t)$, and $ \eta ({\bf x},t) $ includes all the
fluctuations produced by interactions among the unresolved microscopic
degrees of freedom. This noise term clearly influences the dynamics at
mesoscopic scales and therefore the global properties of these
coarse-grained surfaces, in particular their rough or super-rough
nature.  In this work, we mainly focus on the case of white noise of
zero average and amplitude $D$ which is delta-correlated in time as
well as in space
\begin{eqnarray}
\left \langle \eta ({\bf x},t) \right \rangle &=& 0\nonumber\\
\left \langle
\eta ({\bf x},t) \eta ({\bf x}',t') \right \rangle &=& 2 D \delta
^{d}({\bf x}-{\bf x}')~\delta(t-t').
\label{eq1.2} 
\end{eqnarray} 
In equations (\ref{eq1.2}) and below, the symbol $\left \langle \ldots
\right\rangle $ represents ensemble averages over different
realizations of the noise.

Constraints based on symmetry principles are helpful in reducing the
functional ${\cal F}(h) $ to a few standard prescribed forms,
according to the type of growth process it is expected to model. The KPZ
equation \cite{Kardar86}, which is among the most common ones, reads
\begin{eqnarray} \partial_t h  &=&  \nu  \nabla^2 h
+\frac{\lambda }{2} \left( \nabla h \right)^2
+ \eta
\label{eq1.3}
\end{eqnarray}
where $\nu$ and $\lambda$ denote coupling parameters for the linear
and non-linear terms respectively. This equation contains, in addition
to a linear diffusion term, a non-linear term which takes into account
the local growth normal to the interface \cite{Barabasi95,Krug97}.
In the following, fields are assumed to be periodic of characteristic
length $L=V^{1/d}$ with respect to any spatial directions.

\subsection{Finite difference discretizations}
\label{sec:finite}
\noindent In finite-differences methods, one discretizes space by defining
points ${\bf x}_{{\bf j}} = {\bf j} \Delta x $~(${\bf j}$ being a
set of $d$ integer indices) on a cubic lattice of elementary size
$\Delta x$. In the framework of a one step Euler time discretization,
the KPZ time evolution reads
\begin{eqnarray}
h({\bf x}_{{\bf j}},t+\Delta t)&=& h({\bf x}_{{\bf j}},t)+ \Delta t
\left[ \nu \nabla^2 h+ \frac{\lambda }{2} \left( \nabla h \right)^2
\right]_{{\bf x_j}} \nonumber \\
&&+ \Delta t ~\eta\left({\bf x_{\bf j}},t\right),
\label{eq1.4}
\end{eqnarray}
where one  sets
\begin{eqnarray}
\eta({\bf x}_{{\bf j}},t)&=& \sqrt{\frac{24 D}{(\Delta x)^d \Delta t}}
\xi({{\bf j}},t).
\label{eq1.5}
\end{eqnarray}
Here the factor $\xi$ is a noise of zero average and correlation
\begin{eqnarray}
\left \langle \xi ({{\bf j}},t) \xi ({\bf j'},t') \right \rangle &=& 2
\delta_{{\bf j},{\bf j'}}~\delta_{t,t'},
\label{eq1.6}
\end{eqnarray}
and it is taken from a uniform distribution between $-1/2$ and $1/2$.
The prefactor $ \sqrt{\frac{24 D}{(\Delta x)^d \Delta t}}$ ensures
that the noise has a second moment identical to that of the Gaussian
noise integrated over a time interval $\Delta t$ \cite{Risken89}. Note
that the use of a uniform distribution makes no difference and speeds
up the computation as it is well accepted \cite{Amar90}. In the
operators containing spatial derivatives in (\ref{eq1.4}) one must now
introduce their finite-differences approximations. The Laplacian
operator reads
\begin{eqnarray}
\nabla^2 h \left( {\bf x}\right) &=& \frac{1}{\Delta x^2}
\sum_{\mu=1}^{d} [ h({\bf x} + \Delta x~ \widehat{e}_{\mu},t)+\nonumber\\
&&h({\bf x}- \Delta x~ \widehat{e}_{\mu},t) -
 2 h({\bf x},t) ],
\label{eq1.7}
\end{eqnarray}
Where $\widehat{e}_{\mu}$ stands for the basis vector along the
$\mu$-th direction.  For the non-linear term, different possibilities
exist which have been argued to lead to different results
\cite{Lam98,Giacometti01}. We restrict ourselves to the two following
cases: the standard one
\begin{eqnarray}
\left(\nabla h\right)^2 \left( {\bf x}\right) &=& \frac{1}{4 \Delta x^2}
\sum_{\mu=1}^{d} [ h({\bf x} + \Delta x~ \widehat{e}_{\mu},t)-\nonumber\\
&& h({\bf x}- \Delta x~ \widehat{e}_{\mu},t) ],
\label{eq1.8}
\end{eqnarray}
henceforth referred to as (ST), and the one proposed by Lam and Shin
\cite{Lam98}, henceforth referred to as (LS) \cite{note1}
\begin{widetext}
\begin{eqnarray}
\left(\nabla h\right)^2 \left( {\bf x}\right) &=& \frac{1}{3 \Delta
x^2} \sum_{\mu=1}^{d} \{ \left[h({\bf x} + \Delta x
~\widehat{e}_{\mu},t)- h({\bf x},t) \right]^2 + \left[h({\bf x} ,t)-
h({\bf x}- \Delta x~ \widehat{e}_{\mu},t) \right]^2 \\ \nonumber &&
+\left[h({\bf x} + \Delta x ~\widehat{e}_{\mu},t)- h({\bf x},t)
\right] \left[h({\bf x} ,t)- h({\bf x}- \Delta x ~\widehat{e}_{\mu},t)
\right] \}.
\label{eq1.9}
\end{eqnarray}\end{widetext}
It is worth noting here that the LS discretization does not violate a
fluctuation-dissipation theorem which is peculiar of the $(1+1)$-
dimensions, unlike the ST discretization \cite{Lam98,Giacometti01}.

\subsection{Pseudo-spectral discretization}
\label{sec:pseudo-spectral}
\noindent In spectral discretization, the continuum periodic field
$h({\bf x},t)$ is first expanded in Fourier modes $\widehat{h}_{{\bf
q}}(t)$
\begin{eqnarray}
h({\bf x},t) &=& \frac{1}{V} \sum_{{\bf q}} \widehat{h}_{{\bf q}}(t)~
e^{\mathrm{i}{\bf q} {\bf x}},
\label{eq1.10}
\end{eqnarray}
where
\begin{eqnarray}
\widehat{h}_{{\bf q}}(t) &=& \int_{V} d^{d}{\bf x}~h({\bf
x},t)~e^{\mathrm{-i}{\bf q} {\bf x}},
\label{eq1.11}
\end{eqnarray}
and ${\bf q} \equiv (2\pi n_1/L,\ldots,2\pi n_d/L)$ is a wave vector
defined by $d$ integers $n_{\mu}$,~ $\mu=1,\ldots,d$. Similarly, one
expands the noise term $\eta({\bf x},t)$ and thereafter applies the
Fourier transform to the continuum equation (\ref{eq1.3}). An
infinite system of coupled Langevin equations is thus obtained
\begin{eqnarray}
&&\frac{d \widehat{h}_{{\bf q}(t)}}{dt}= -\nu {\bf q}^2 ~
\widehat{h}_{{\bf q}}(t)-\nonumber\\
&& \frac{\lambda}{2V} \sum_{{\bf k},{\bf
k^{\prime}}} ({\bf k} \cdot {\bf k^{\prime}} ) \widehat{h}_{{\bf
k}}(t) \widehat{h}_{{\bf k^{\prime}}}(t) \delta_{{\bf k}+{\bf
k^{\prime}},{\bf q}} + \widehat{\eta}_{{\bf q}}(t).
\label{eq1.12}
\end{eqnarray}
The Fourier modes $\widehat{\eta}_{{\bf q}}$ satisfy, according to
equations (\ref{eq1.2}), the correlation:
\begin{eqnarray} 
\left \langle \widehat{\eta}_{{\bf q}}(t) \widehat{\eta}_{{\bf
q^{\prime}}}(t') \right \rangle &=&2 V D \delta_{{\bf q},-{\bf
q^{\prime}}} \delta(t-t').
\label{eq1.13}
\end{eqnarray}

The spectral discretization then consists in projecting the infinite
system (\ref{eq1.12}) on the vector space of periodic functions of
period $L$ with only a finite number of non-vanishing Fourier modes.
To be more specific, it is assumed that wavevectors ${\bf q}$, ${\bf
k}$ and ${\bf k^{\prime}}$ in (\ref{eq1.12}) belong to a set ${\cal
S}$ indexed by integers $ |n_{\mu}| \le {N/2}$ for $\mu=1,\ldots,d$.
In this approximation, the dynamical equations conserve their original
form with finite sums replacing the original infinite ones.  The noise
term and the spatial derivatives are discretized in a similar way.
This approach is thus more consistent when compared to
finite-differences.  Furthermore, it provides exact results for the
Edward Wilkinson equation (EW), which can be read off from the KPZ
equation (\ref{eq1.3}) with $\lambda =0$.  In this case the KPZ
equation is reduced to a set of $(N+1)^d$ uncoupled complex Langevin
equations. With our method the regularization is performed in Fourier
space and its efficiency in reproducing the features of the continuum
equation should be compared to that of finite differences at constant
number of Langevin equations. Note that the lattice spacing is such
that $a = L/N$.
 
The temporal discretization for the above complex equations is performed
by an Euler one time step method \cite{note2}.  In this case,
computation of the linear term as well as the noise is straightforward
at each time step. On the other hand, the computation of the non-linear term
\begin{eqnarray}
\widehat{\chi_q}&=& -\frac{1}{2V}\sum_{{\bf k},{\bf q}- {\bf k} \in
{\cal S} } \left[{\bf k}\cdot ({\bf q}- {\bf k}) \right]
\widehat{h}_{{\bf k}}(t) \widehat{h}_{{\bf q}- {\bf k}}(t),
\label{eq1.14}
\end{eqnarray}
which originates from one-half of the square of gradient $\nabla
h({\bf x},t)$, necessitates more care. We use a pseudo-spectral
algorithm which hinges on the following four steps:
\begin{description}
\item[(a)] Extension of the momentum space in such a way to have $M
+1$ modes ($M>N$) per direction instead of $N+1 $ modes.  The
supplementary Fourier modes $\widehat{h}_{{\bf q}}$ are simply set to
zero.
\item[(b)] Computation, in this extended momentum space, of the
spectrum of gradient $\nabla h({\bf x},t)$ by sheer multiplication $
{\bf q}~ \widehat{h}_{{\bf q}}$.
\item[(c)] Computation by Fourier transform of $\nabla h({\bf x},t)$
in real space at $(M+1)^d$ collocations points ${\bf x}_{\bf j}$
located on a d-dimensional hyper-cubic lattice of elementary size
$L/M$. This is followed by the straightforward calculation of
$1/2\left [ \nabla h({\bf x},t)\right]^2$ at these same collocations
points.
\item[(d)] By a further Fourier transform of these $(M+1)^d$ values of
$ 1/2 \left[ \nabla h({\bf x},t)\right]^2$, one gets the correct sums
$\widehat{\chi_q}$ for all wavevectors ${\bf q} \in {\cal S} $ if $M$
is sufficiently large. The values found for external supplementary
Fourier modes are simply discarded.
\end{description}
The entire procedure can be efficiently implemented using a standard
Fast-Fourier package \cite{fftw}. Without a preventive
extension of the number of Fourier modes for this specific
computation, the exact non-linear term $\widehat{\chi_q}$ would not be
obtained for $ {\bf q}\in {\cal S}$: this is the well-known aliasing
problem \cite{Canuto88,Numericrecipes}. The minimum (and hence most
efficient) choice for $M$ turns out to be $3N/2$ \cite{Canuto88}. The
reasons for this are detailed in Appendix \ref{sec:appendixa}.

\section{Some relevant quantities.}
\label{sec:relevant}
\noindent
Kinetically rough surfaces generated by stochastic differential
equations such as (\ref{eq1.3}) generally possess scale invariant
properties. We recall below some characteristic quantities in which
this behavior becomes manifest as a power law, and define the relevant
critical exponents.

The correlation between fluctuations of height between any two points
${\bf x}$, ${\bf x}+{\bf r}$ located at distance $r$, is given by the
height-to-height correlation function
\begin{eqnarray}
G_n(r,t) &=& \left ( \frac{1}{V} \int_V d^{d}{\bf x} \left \langle|
h({\bf x},t)-h({\bf x}+{\bf r}, t)| ^n \right \rangle \right)^{1/n}
\label{eq3.1}
\end{eqnarray}
where $n$ is, in general, any real positive number. It is generally
assumed that $ G_2(r,t)$ has the scaling form
\begin{eqnarray}
G_2(r,t) &=& r^{\chi} M\left(\frac{t}{r^z} \right),
\label{eq3.2}
\end{eqnarray}
where the function $M(y)$ is such that
\begin{equation}
M(y)\sim\left\{
\begin{array}{lll}
 \mathrm{const} & & y \to \infty \\
y^{\beta} &  & y \to 0
\end{array}
\right. \qquad
\label{eq3.3}
\end{equation}
with $ \beta \equiv \chi/z $. $z$ is the dynamical exponent, and
$\chi$ is the roughness exponent. Correlations are then of the form
$G_2(r) \sim r^{\chi} $ in the limit $r \ll t^{1/z}$, whereas they
follow the asymptotic limit $G_2(r,t) \sim t^{\beta}$ for $r \gg
t^{1/z}$.  In the absence of multiscaling \cite{Mohayaee01,Lopez99},
the same asymptotic scaling behaviors should be clearly
envisaged for different $n$'s. The law (\ref{eq3.2}) can be also
connected to the scaling of the roughness $W(t,L)$. This latter
quantity, which measures fluctuations amplitudes, is defined as
\begin{eqnarray}
W^2(t,L) \equiv  \langle \left( h - \overline{h}_L\right)^2 \rangle,
\label{eq3.4}
\end{eqnarray}
where $\overline{h}_L$ denotes the average height over a volume
$V=L^d$. For the Family-Vicsek ansatz \cite{Family90}, such a global
scaling reads
\begin{eqnarray}
W(t,L) &=& L^{\chi} N\left( \frac{t}{L^z} \right),
\label{eq3.5}
\end{eqnarray}
where function $N(y)$ is such that $N(y) \to \mathrm{const}$ when $y
\to \infty $ and $N(y) \sim y^\beta $ when $y \to 0$.  The relation
between $W(t,L)$ and $G_2(r,t)$ is discussed in Appendix
\ref{sec:appendixb}. The exponents are similar to those
of (\ref{eq3.3}): quantity $\beta $ is related to the short time
dynamics $W(t,L) \sim t^{\beta}$ and $\chi $ to the asymptotic
saturation width. $W(t,L) \sim L^{\chi}$. Moreover, note that the
roughness may be expressed as (see Appendix \ref{sec:appendixc})
\begin{eqnarray}
W^2(t,L)&=& \frac{1}{V} \sum_{q \ne 0} S(q,t),
\label{eq3.6}
\end{eqnarray}
where $S(q,t)$ denotes the structure factor
\begin{eqnarray}
S(q,t)&=& \frac{1}{V} \left \langle \hat{h}_q(t) \hat{h}_{-q}(t)
\right \rangle.
\label{eq3.7}
\end{eqnarray}
In momentum space, one infers the scaling behavior to be (see Appendix
\ref{sec:appendixc})
\begin{eqnarray}
S(q,t) &=& q^{-(d+2 \chi)} \Phi(q^z t),
\label{eq3.8}
\end{eqnarray}
where the function $\Phi$ is such that
\begin{eqnarray}
\Phi\left( y \right)\sim
\left\{
\matrix{
\mbox{const} \qquad \mbox{for } y \gg 1,  \cr
y^{d+2\chi} \qquad \mbox{for } y \ll 1.
\cr}
\right.
\end{eqnarray}
Consequently $S(q,t) \sim q^{-(d+2 \chi)}$ for large $t$ and $ S(q,t)
\sim t^{2 \beta+d/z}$ for small $t$.

In our simulations we computed the skewness parameter $s$, which
is the (scaled) third moment of the height fluctuations
\begin{eqnarray}
s(t)= \frac{\left\langle \left( h - \overline{h}_L \right)^3
\right\rangle } {\left\langle \left( h - \overline{h}_L \right)^2
\right\rangle ^{3/2} }.
\label{eq3.9}
\end{eqnarray}
As observed by Krug et al.~\cite{Krug92} for $d=1$, this parameter is
a good measure of the asymmetry of the distribution of height
fluctuations.

Finally, we recall that the KPZ equation carries a close analogy with
turbulence theory. For instance, the equivalent of a Reynolds number
for the KPZ equation can be defined as \cite{Hentschel91}
\begin{eqnarray}
\epsilon &=& \frac{D \lambda^d}{\nu^{d+1}}
\label{eq1.19}
\end{eqnarray}
Moreover, exact heuristic arguments show the existence of the
following two length scales
\begin{eqnarray}
\l_{\mathrm{in}} &=& \left(\frac{\nu^{d+3}}{ D \lambda^{d+2}}\right)
^{1/2}
\label{eq1.20}
\end{eqnarray}
\begin{eqnarray}
\l_{\mathrm{out}} &=& \left(\frac{D}{ \lambda}\right) ^{1/(d+1)}.
\label{eq1.21}
\end{eqnarray}
For a length $l$ smaller than $l_{\mathrm{in}}$, the diffusion linear
term is predominant as in the case of viscous scales observed in
turbulent flows. Conversely, the nonlinear term is dominant for length
$l \gg l_{\mathrm{in}}$, as for the inertial scales in turbulent
flows. In fluid dynamics, the outer scale is given by the largest
scale available, which is generally provided by the geometry of the
specific problem. In the KPZ case \cite{Hentschel91}, it is given by
imposing that for scales $l_{\mathrm{in}} \le l \le
l_{\mathrm{out}}$, the surface is rough. Note that, since the
dimensionless number $\epsilon$ is given by
\begin{eqnarray}
\epsilon &=& \left(\frac{l_{out}}{l_{in} }\right)^{2(d+1)/(d+3)},
\label{eq1.22}
\end{eqnarray}
there exists an equivalent of the inertial range for large equivalent
Reynolds numbers $\epsilon \gg 1 $. In this regime, the typical
fluctuation of scale $l$ is of amplitude $h_l$ and varies with a
typical time $t_l$ given by
\begin{eqnarray}
h_l \equiv l^{2/(d+3)} \left(\frac{D}{ \lambda}\right) ^{1/(d+3)}
\qquad t_l \equiv \frac{l^{2}}{ \lambda h_l }.
\label{eq1.23}
\end{eqnarray}
In the context of the KPZ equation, the inertial range is represented
by the strong-coupling regime. This corresponds to $\epsilon \gg 1$.
However, the requirement of the existence of a non-negligible inertial
region, that is $l_{\mathrm{out}} \gg l_{\mathrm{in}}$, yields
$\epsilon^{f(d)} \gg 1$ with $f(d)=(d+3)/2(d+1)$. Since $f(d)<1$ for
$d>1$ then $\epsilon^{f(d)}$ is a slowly increasing function of
$\epsilon$. The above remark explains the difficulties of the
numerical simulations to reach unambiguous results, due to the long
transients which are present in $(2+1)$ and higher dimensions, as we shall
further discuss below.
 
\section{A test case: $(1+1)-$ dimensions }
\label{sec:case1}
\begin{figure}
\includegraphics[width=8cm,clip]{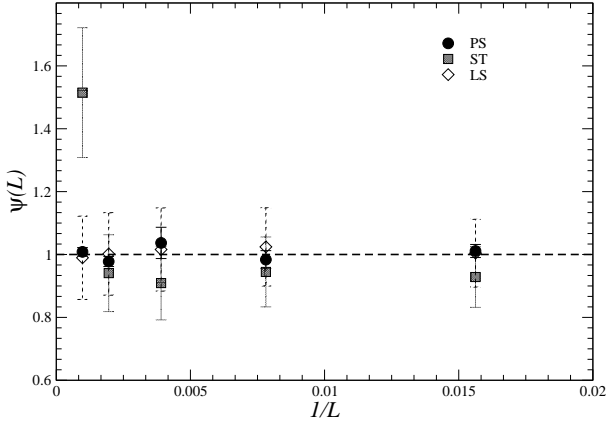}
\caption{Quantity $\psi(L)\equiv \sqrt{12 \nu / D L} W(L)$ in
$(1+1)$-dimensions, computed as a function of the size $L$.  $W(L)$ is
the steady-state roughness as computed using the standard finite
difference discretization (ST) (with $\Delta x=1$), the Lam-Shin
finite difference discretization (LS) (with $\Delta x=1$), and the
pseudo-spectral discretizations (PS) with $L=N$. All these
discretizations have an identical number of degrees of freedom. The
number of configurations used in the average is also identical in the
three cases. The error bars have been distinguished for clarity: 
grey line for ST, dashed line for LS, solid line (barely visible) for PS.}
\label{fig1}
\end{figure}

\noindent
We compare the performance of the various method in the
$(1+1)$-dimensional case. In this instance, the exact value for the
steady-state roughness of the continuum equation is known to be
\cite{Krug92}
\begin{eqnarray}
W(L)&=& \sqrt{\frac{D}{12 \nu}}~L^{1/2}.
\label{eq3.10}
\end{eqnarray}
Such a quantity $W(L)$ has been computed for sizes up to $L=1024$ and
parameters $\nu=1$, $\lambda=3$, $D=1$, and averaged over $100$
different realization of the noise and many different
steady-state configurations. We have used (a) the finite-differences
scheme given by Eq.~(\ref{eq1.8}) (ST), (b) the finite-differences
scheme given by Eq.~(\ref{eq1.9}) (LS) and (c) the pseudospectral
scheme (PS). For finite-differences we have set $\Delta x =1$
corresponding to $L=N$ for the pseudo-spectral method.  
A first comparison is reported in Fig.~\ref{fig1}, 
where the quantity $\psi(L)=
\sqrt{12 \nu/ D L}~W(L)$ is plotted for the three cases. 
We note that error bars refer to 
fluctuations within the $100$ different noise configurations and that
the considered three cases are compared within {\it identical} statistics.

Unlike the standard discretization, both the
pseudo-spectral and the Lam-Shin discretizations yields exact
values for the amplitude (dashed line in Fig.~\ref{fig1}), 
within the error bars. However the pseudo-spectral method yields much
smaller fluctuations compared to the Lam-Shin discretization as
it can be inferred by comparing the error bars for LS (dashed lines)
with those of the PS method (solid line) which are barely visible
in Fig.\ref{fig1} being of the order of the points size. 

\begin{figure}
\includegraphics[width=8cm,clip]{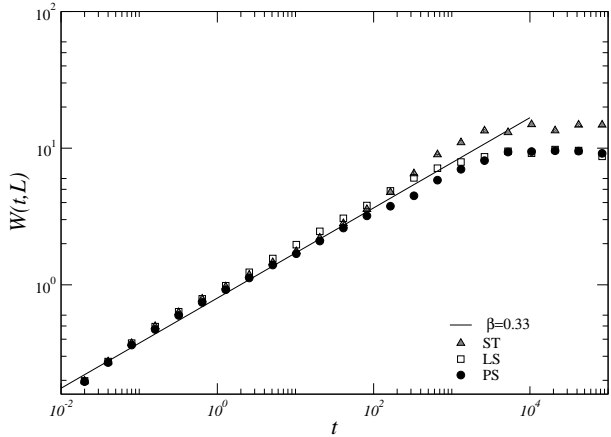}
\caption{ Roughness $W(t,L)$ in the $(1+1)$-dimensional case as a
function of time $t$ for $\lambda=3$. Here the lateral size is
$L=1024$, the number of configurations is $50$, and ST, LS and PS
stands for standard, Lam-Shin, and pseudo-spectral discretizations
respectively. The solid line has a slope corresponding to $\beta=0.33$.}
\label{fig2}
\end{figure}

As a further support to this
result, we compute the time evolution of roughness before saturation,
which allows to obtain the critical exponent $\beta$ whose exact value
is $1/3$. In this case the roughness is averaged over $50$ different
configurations in all cases. Fig.~\ref{fig2} depicts the
result. The pseudo-spectral
method follows rather accurately the exact value of the exponent for
at least $3$ decades, whereas the LS method slightly overestimates
that value. For the pseudo-spectral discretization, we have also computed (not
shown) the roughness exponent $\chi$ with the two different methods
illustrated in the next section, always finding an excellent
agreement with the exact value $\chi=1/2$.

\section{Case $(2+1)-$ dimensions}
\label{sec:case2}
\noindent
Let us consider the most physically relevant case i.e.~that of
bi-dimensional substrate $d=2$. To this purpose, results for various
finite differences discretizations and for the pseudospectral
discretization are provided for sizes up to $L=256$ and $L=512$
respectively. Computations are averaged over a number of
configurations typically of the order of $50-100$. The values of $\nu$
and $D$ are both set to $1$ throughout the rest of this section.

\subsection{Finite difference}
\label{subsec:finite}
\begin{figure}
\includegraphics[width=8cm,clip]{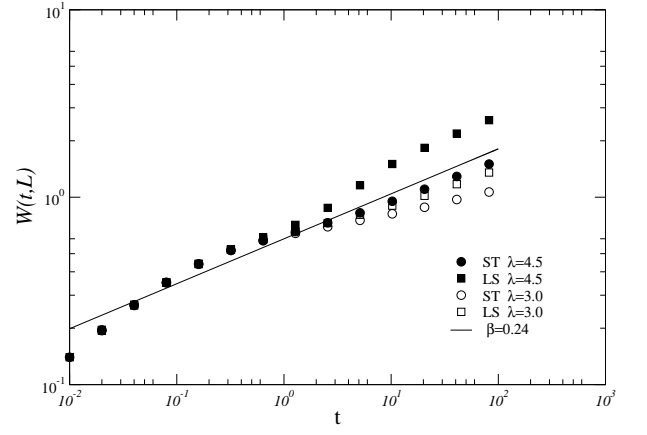}
\caption{Roughness $W(t,L)$ obtained from a finite-difference scheme
in $(2+1)$-dimensions, as a function of time $t$ for $\lambda=3,4.5$.
Here the lateral size is $L=256$, the number of configurations is
$50$, and ST and LS stands for Standard and Lam-Shin discretizations
respectively. The solid line has a slope corresponding to $\beta=0.24$.}
\label{fig3}
\end{figure}

\noindent
To the best of our knowledge, no papers have so far considered a
comparison among various finite differences discretizations in
dimensions $d=2$. As previously mentioned, we envisage two relevant
finite-differences approximations (ST) and (LS) of the KPZ
equation. In Fig.~\ref{fig3}, a comparison of the roughness $W(t,L)$
as a function of time $t$ is reported for these two methods and two
values of the non-linear parameter ($\lambda=3$ and
$\lambda=4.5$). From the figure, one can appreciate the presence of
the three regimes (linear regime, KPZ regime and saturation regime)
which are characteristic of any numerical solution of the KPZ equation
in $(2+1)$-dimensions when starting from a flat configuration. Both
the above values of $\lambda$ yield a non-negligible region of KPZ
regime. From Eq.~(\ref{eq1.19}), one has $\epsilon=\lambda^2$ and
$\epsilon^{f(d)}=\lambda^{10/6}$, which for $\lambda=3$ give
$\epsilon=9$ and $\epsilon^{f(d)}=6.240\ldots$ respectively. For such
a moderate non-linearity ($\lambda=3$) the inertial region is already
present, and it increases as $\lambda$ increases. However, an
instability already noticed in previous simulations
\cite{Amar90,Moser91} is present for larger value of $\lambda$. The
curvature present in the $\lambda=4.5$ case of Fig.~\ref{fig2}, stems
from this instability, and it is probably associated with an
overestimation of the nonlinearity.

Our results for the critical exponents are in good agreement with
previous results reported in the literature for finite-difference schemes. 
A summary of all these results is provided in Table \ref{table1} for
completeness \cite{note4}.

\subsection{Pseudo-spectral method}
\label{subsec:pseudo1}
\begin{figure}
\includegraphics[width=8cm,clip]{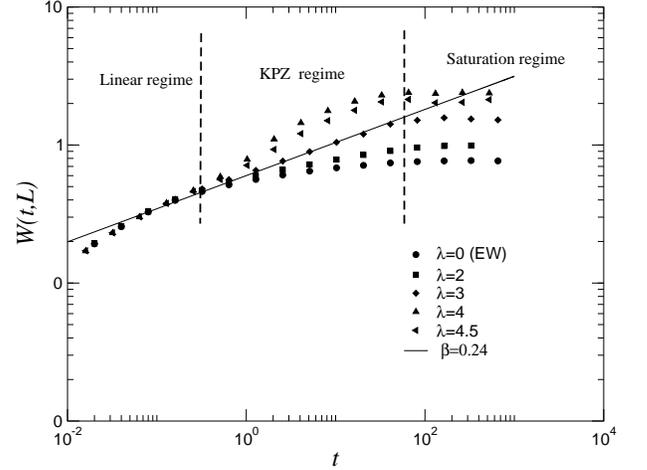}
\caption{Roughness $W(t,L)$ as a function of time $t$ for various
values of $\lambda$ as obtained from the pseudo-spectral methods. Here
the lateral size is $L=128$ and the average is over $100$ different
configurations. We have schematically indicated the typical three regimes
found in the simulations.}
\label{fig4}
\end{figure}

\begin{figure}
\includegraphics[width=8cm,clip]{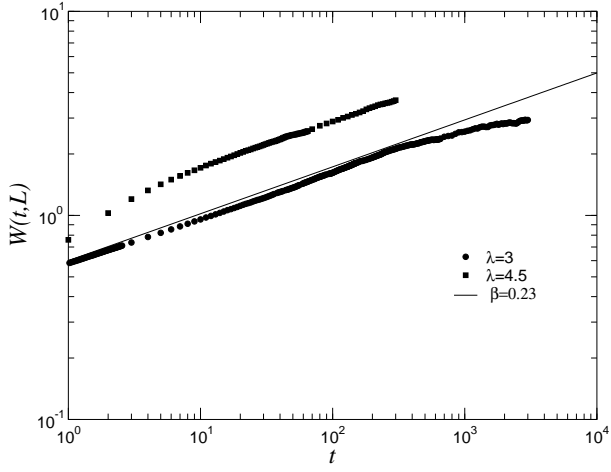}
\caption{$W(t,L)$ in $(2+1)$-dimensions, as a function of $t$ 
when $\lambda=4.5$ and
$\lambda=3$. The lateral size is $L=512$ and the number of
configurations is $17$. The dashed line has a slope corresponding to
$\beta=0.23$.}
\label{fig5}
\end{figure}

\noindent 
The roughness (\ref{eq3.6}) computed by the pseudo-spectral
method is depicted in Fig.~\ref{fig4} for $L=128$, and various values
of $\lambda$.  It has been computed after averaging over many
realizations of noise (typically of the order of $50-100$).  The value
$\lambda=3$ numerically corresponds to the optimal choice used in
previous works \cite{Amar90,Guo90,Moser91,Chakrabarti91,Beccaria94} in
which the strong coupling regime is well displayed (see
Fig.~\ref{fig4}).  The power law for times $t \le L^z$ has been
reassessed (see Fig.~\ref{fig5}) and exponent $\beta$ is evaluated to
be $\beta=0.229 \pm 0.05$ for $\lambda=3$ and $4.5$ ($L=512$ is the
size used to get this result).

\begin{figure}
\includegraphics[width=8cm,clip]{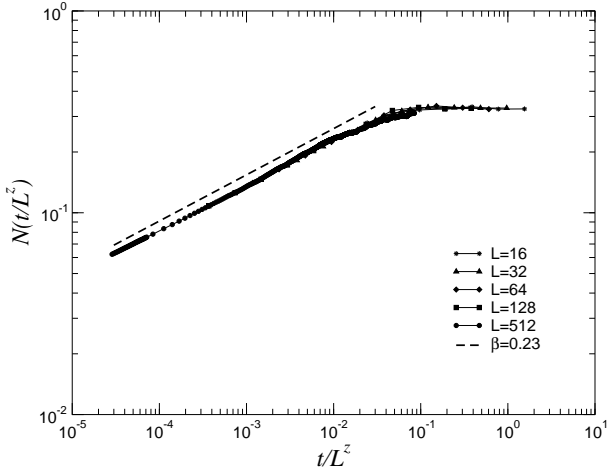}
\caption{Collapse plot of the universal function of $N \left( t/L^z
\right)$ when $\lambda=3$ and various sizes $L$. The dashed line
corresponds to $\beta=0.23$. Parameters are identical to
Fig.~\ref{fig4}.
The obtained values for the exponents are
$\chi=0.37 \pm 0.02 $ and $z=1.67 \pm0.05$.}
\label{fig6}
\end{figure}

The universal function $N \left( t/L^z \right)=W\left( t/L^z
\right)/L^\chi$  is plotted for $\lambda=3$ in
Fig.~\ref{fig6}. The exponents $z$ and $\chi$ can be estimated
using a recent method devised in Ref.~\cite{Bhattacharjee01}, which
allows the calculation of the two exponents simultaneously with good
accuracy.
This method provides a more accurate estimate of the roughness 
exponent $\chi$ with respect to 
the more commonly used procedure of computing the
saturation value of $W$ at different system sizes. We obtain $\chi
=0.37 \pm 0.02 $ and $z=1.67 \pm0.05$.

\begin{figure}
\includegraphics[width=8cm,clip]{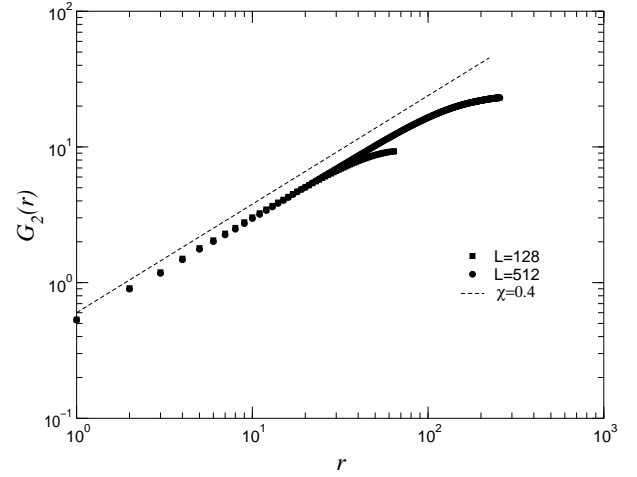}
\caption{\label{fig7} Height-height correlation function $G_2(r)$ for
$\lambda=3$ and sizes $L=128$, $L=512$. The dashed line corresponds to
$\chi=0.4$.}
\end{figure}

\begin{figure}
\includegraphics[width=8cm,clip]{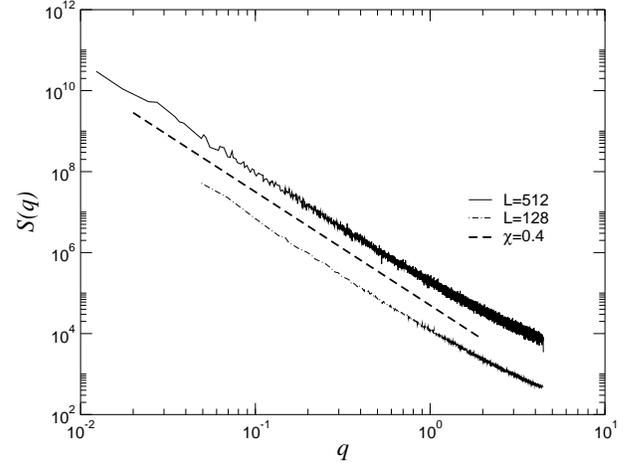}
\caption{\label{fig8} Structure factor $S(q)$ for $\lambda=3$ and
sizes $L=128$,$512$. The dashed line corresponds to $\chi=0.4$.}
\end{figure}

Two alternative procedures to independently calculate the roughness
exponent is based on the computation of the height-height correlation
function $G_2(r)$ and of the structure factor $S(q)$ at stationarity.
Its results are reported in Figs.~\ref{fig7} and \ref{fig8}. Our best
values for the roughness exponent $\chi$ are $\chi =0.38 \pm 0.02 $
from the correlation function and $\chi=0.40 \pm 0.01$ from the
structure factor. Both these values are, within the error bars,
compatible with previous numerical results on the continuum KPZ
equation in real space
\cite{Amar90,Guo90,Moser91,Chakrabarti91,Beccaria94} and with recent
extensive simulations on RSOS models \cite{Marinari00}. As explained
in Ref.~\cite{Siegert96}, the computation of the structure factor
yields more reliable results of the exponent $\chi$ with respect to
the commonly used procedure of extracting it from the value of the
saturation roughness at various sizes.

\begin{figure}
\includegraphics[width=8cm,clip]{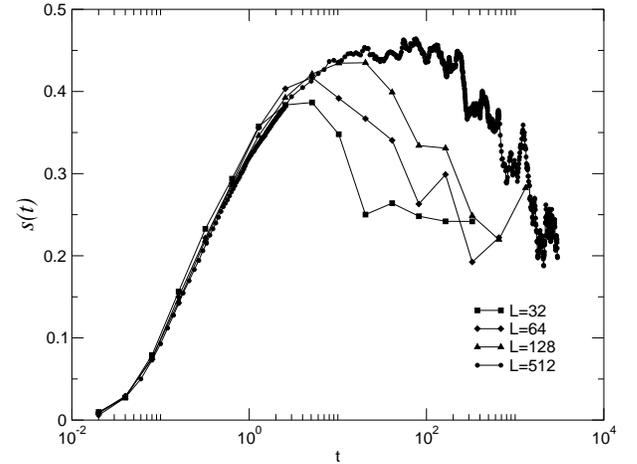}
\caption{\label{fig9} Skewness $s(t)$ for $\lambda=3$ and sizes $L$
ranging from $L=32$ to $L=512$.}
\end{figure}

Fig.~\ref{fig9} depicts the time evolution of skewness $s(t)$ for
different sizes $L$. For $\lambda >0$, the quantity $s(t)$ reaches an
asymptotic value ($0.25-0.30$) indicating a clear-cut asymmetry as
opposed to the EW case ($\lambda=0$). This asymmetry in $(2+1)$ is
intrinsically different from the one observed for short times both on
the KS $(1+1)$ \cite{Sneppen92} and for the single-step
one-dimensional model \cite{Krug92}, since in that instance the
asymptotic value of the skewness should be zero in both cases, due to
the Gaussian nature of the probability distribution function. The
existence of a well defined peak for $s(t)$ suggests that this
quantity could be a better indicator of the onset of the KPZ regime
with respect to the roughness.

\begin{figure}
\includegraphics[width=8cm,clip]{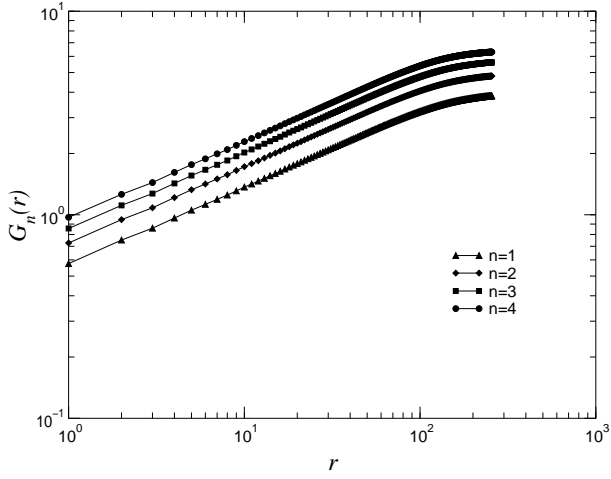}
\caption{\label{fig10} Various moments of the height difference
distribution at stationarity $G_n(r)$ for $\lambda=3$ and size
$L=512$. Note that the largest value of $r$ corresponds to $L/2$.}
\end{figure}

\begin{figure}
\includegraphics[width=8cm,clip]{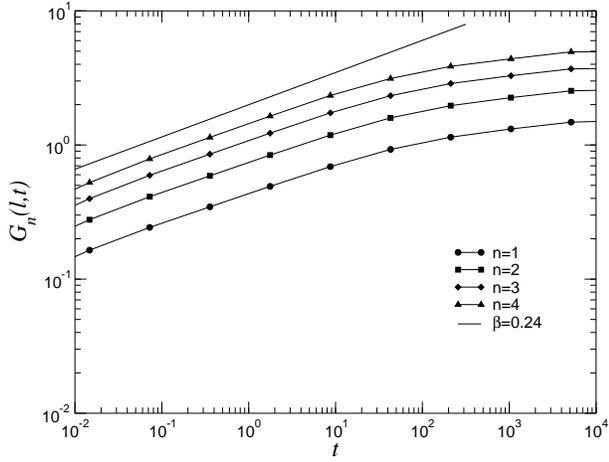}
\caption{Various moments of the height-height correlation
 functions $G_n(l,t)$ for a
fixed $l \sim 6 L/N$ and $\lambda=3$, $L=512$. The solid line
corresponds to $\beta=0.24$.}
\label{fig11}
\end{figure}

It has been suggested \cite{Krug94,Lopez99,Mohayaee01}, that possible
multiscaling behaviors can be inferred by computing the {\it local}
fluctuations $ G_n(l,t)$ (where $L/N < l < L/2$, and the extreme case
in which $l$ equals the lattice spacing $L/N$ corresponds to the
average values of powers of the surface gradient) with respect to
time. The reason for this can be traced back to the fine balance
between different terms present in the Langevin equation, necessary to
get standard scaling. On that basis, one does not expect multiscaling
in the KPZ equation case \cite{Lopez99}. A direct computation confirms
this. First we note that different moments of the stationary
state $G_n(r) $ depicted in Fig.~\ref{fig10} have identical
scaling. Next we consider the time evolution of $G_n(l,t)$ reported in
Fig.~\ref{fig11}. This type of calculation has been already performed
in Ref.~\cite{Krug94} and \cite{Mohayaee01} in the context of discrete
models by considering the fluctuations of nearest-neighbors points as
a function of time. Here calculations are performed for points located
few lattice spacings away i.e.~$l$ is included within a
circular region $5 a \le l \le 8 a $ \cite{note3}. It is apparent that
different moments behave in an equal way in agreement with scaling
arguments based on a Flory theory \cite{Lopez99}.

As a remark, we note that it would be very interesting to compare the
above result with an analogous calculation for the
Kuramoto-Shivashinsky equation (KS), since there are both analytical
\cite{Yakhot81,Jayaprakash93} and numerical
\cite{Zaleski89,Sneppen92,Boghosian99} evidences that the (KS) can be
mapped, at a coarse grained level, onto a KPZ with a large surface
tension coupling $\nu$.

\section{Conclusions}
\label{sec:conclusions}
\noindent
We have presented a comprehensive study of the pseudo-spectral method
applied to the numerical solution of the KPZ equations. The method
can be reckoned as an alternative and highly reliable procedure to the
widely exploited finite-difference schemes, and its use in the context
of stochastic partial differential equation is new. At the price of a moderate
increase in the numerical effort, the method offers an improved
accuracy in the computation of the critical exponents, since it does
not suffer of the major drawbacks which are common to all the
finite-difference schemes in real space. The reason for this is that
the functional form of the continuum equation is guaranteed to be
maintained, the only approximation made in the calculation being the
truncation to a finite rather than infinite number of modes. We have
detailed how one can carefully deal with the computation of the
non-linear mode, by using a back and forth transformation between real
and momentum space which is more efficient than a brute force
computation of the non-linear term directly in momentum space. We have
shown how finite-difference schemes lead to non-negligible
differences in the universal behavior in the temporal region which is
accessible to numerical simulations, and explained why
the pseudo-spectral method is the one that both theoretically and
practically most closely approaches the continuum limit. As a
non-trivial application of the method, we have shown, using extensive
simulations and comparing the various different schemes, that the
currently most accepted results for the critical exponents $\beta$ and
$\chi$ in $(2+1)$-dimensions, as obtained from RSOS simulations, 
are directly accessible. We
have also presented different ways of computing the roughness exponent
$\chi$ which yield rather precise and consistent results.

While our results are confined to the KPZ equation in $(1+1)$ and
$(2+1)$ dimensions, the method is fully general and can be extended
to higher dimensions and other non-linear continuum equation of
the Langevin class.

We believe that the results presented in this work open new
perspectives in the computation of the universality classes of
non-equilibrium problems by avoiding the uncontrolled use of spatial
discretizations from the outset.  

\acknowledgments 
Funding for this work was provided by a joint CNR-CNRS exchange
program (number 8006), MURST and INFM. We thank Flavio Seno and
Somedra Bhattacharjee for the use of their collapse plot code.

\appendix
\section{The aliasing problem}
\label{sec:appendixa}
\noindent 
In this appendix we discuss the necessity of extending the number of
Fourier modes to compute the non-linear terms (\ref{eq1.14}) at each
time step, and explain why $M=3N/2$ is the optimal choice for the size
of the extended space. For simplicity, we will treat the $d=1$ case,
the extension to higher dimension being straightforward.

\begin{figure}
\includegraphics[width=8cm,clip]{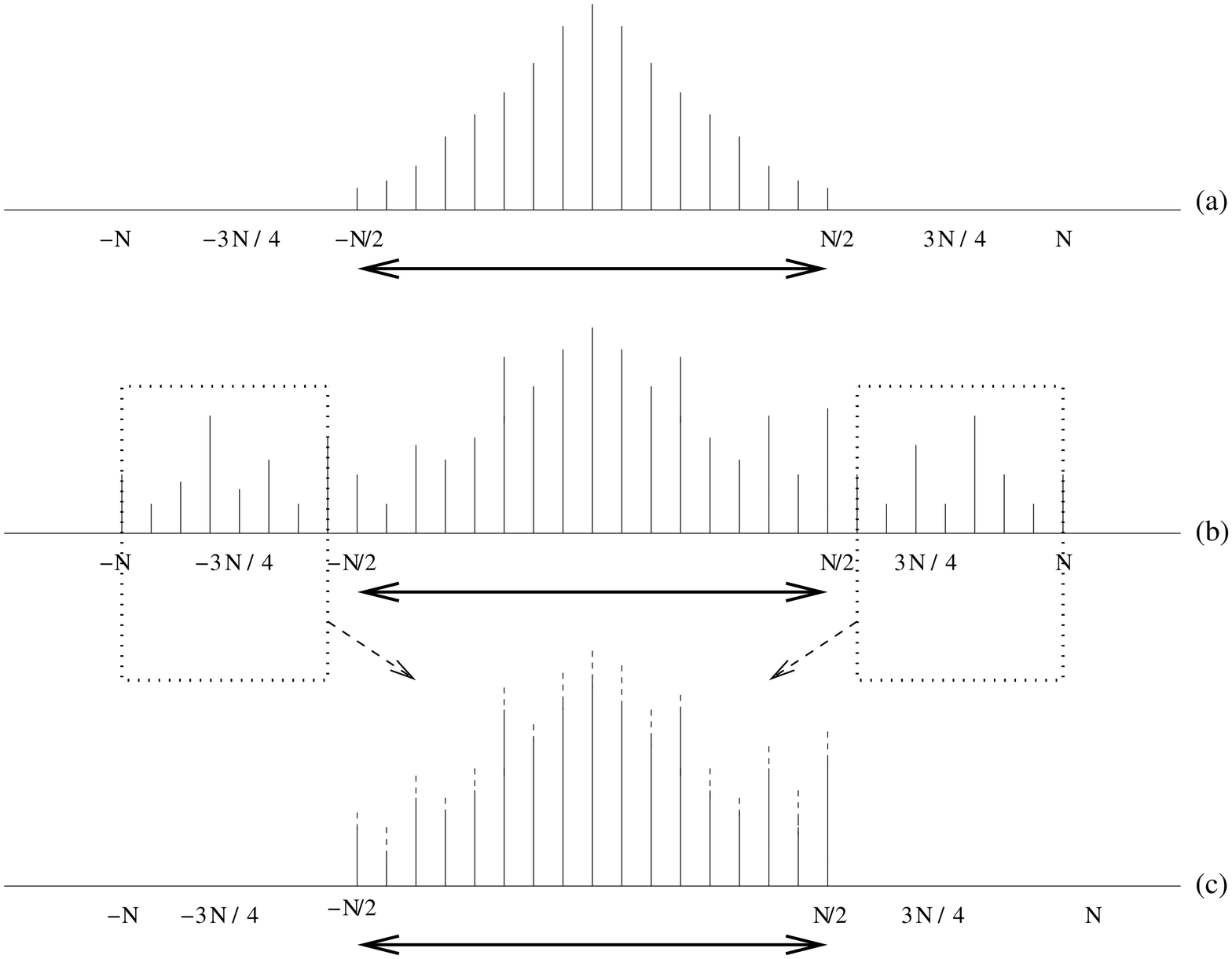}
\caption{Illustrating the aliasing problem when $M=N+1$. The arrow
line indicates the modes used in the computation. (a) the original
field $ h(x,t) $; (b) the true function $ 1/2 \left[ \partial_x
h(x,t)\right]^2$; (c) the truncated function associated to $ 1/2
\left[ \partial_x h(x,t)\right]^2$ when $M=N+1$ }.
\label{fig12}
\end{figure}

Consider the function $h(x,t)$ at time $t$. By definition, it contains
$N+1$ non vanishing modes (see Fig.~\ref{fig12})
\begin{eqnarray}
h(x,t) &=& \frac{1}{L} \sum_{k=-N/2}^{N/2} \widehat{h}_{q_{k}}(t)~
e^{\mathrm{i} { q}_{k} x},
\label{fourdimension1}
\end{eqnarray}
with $q_{k}= 2\pi k/L $. In dimension $d=1$, one has ${\cal S} =
\left\{q_{k}= 2\pi k /L, \;\;N/2 \le k \le N/2 \right\}$ and
$V=L$. The computation of the gradient $\partial_x h(x,t)$ or its
square at any given point, can be carried out correctly by steps
(a)-(c) without any extension of the momentum space. Instead, step (d)
is more delicate and it does require the momentum space extension. To
understand this point, let us tackle the following more general
question. Consider any given periodic function of period $L$
\begin{eqnarray}
u(x,t) &=& \frac{1}{L} \sum_{p=-\infty}^{\infty} \widehat{u}_{{
q}_{p}}(t)~ e^{\mathrm{i} { q}_{p} x}.
\label{fourdimension2}
\end{eqnarray}
Assume we want to approximate $ u(x,t)$ by a truncated function $
u^{tr}$ only containing $2P+1$ non vanishing modes
\begin{eqnarray}
u^{tr}({x},t) &=& \frac{1}{L} \sum_{p=-P }^{P } \widehat{u}_{{
q}_{p}}^{tr}(t)~ e^{\mathrm{i} { q}_{p} x}.
\label{fourdimension3}
\end{eqnarray}
Since such a function $u^{tr}$ is completely defined if it is known for
$M \equiv 2P+1$ points, we choose $ u^{tr}$ by imposing that it
takes the same values as $u$ on $M$ collocation points $x_{j}= j L/ M$ $j=0,
\cdots, M-1$. What is the error made on the Fourier modes or,
equivalently, how different are $\widehat{u}_{{ q}_{p}}$ and
$\widehat{u}_{{ q}_{p}}^{tr}$ for $-N/2 \le p \le N/2$?

For collocation points $x_{j}= L j/M$ $j=0, \cdots, M-1$, the
following relation holds
\begin{eqnarray}
\sum_{p=-\infty}^{\infty} \widehat{u}_{ q_{p}}(t)~ e^{\mathrm{i} {
q}_{p} x_{j}} = \sum_{p=-P}^{P} \left ( \sum_{r=-\infty}^{\infty}
\widehat{u}_{ q_{p+rM}} \right )~ e^{ \mathrm{i} q_{p} x_{j}}.
\label{fourdimension4}
\end{eqnarray}
By identification of (\ref{fourdimension3}) and
(\ref{fourdimension4}), one gets
\begin{eqnarray}
\widehat{u}_{{ q}_{k}}^{tr}(t) = \sum_{r=-\infty}^{\infty}
\widehat{u}_{{ q}_{k+rM}} \label{fourdimension5}
\end{eqnarray}
for $ - P \le k \le P$. Going back to the momentum space, the true
spectrum is thus somewhat modified: this is known as the aliasing
problem. Note that large wavenumbers are, in principle, more affected
than small wavenumbers.

\begin{figure}
\includegraphics[width=8cm,clip]{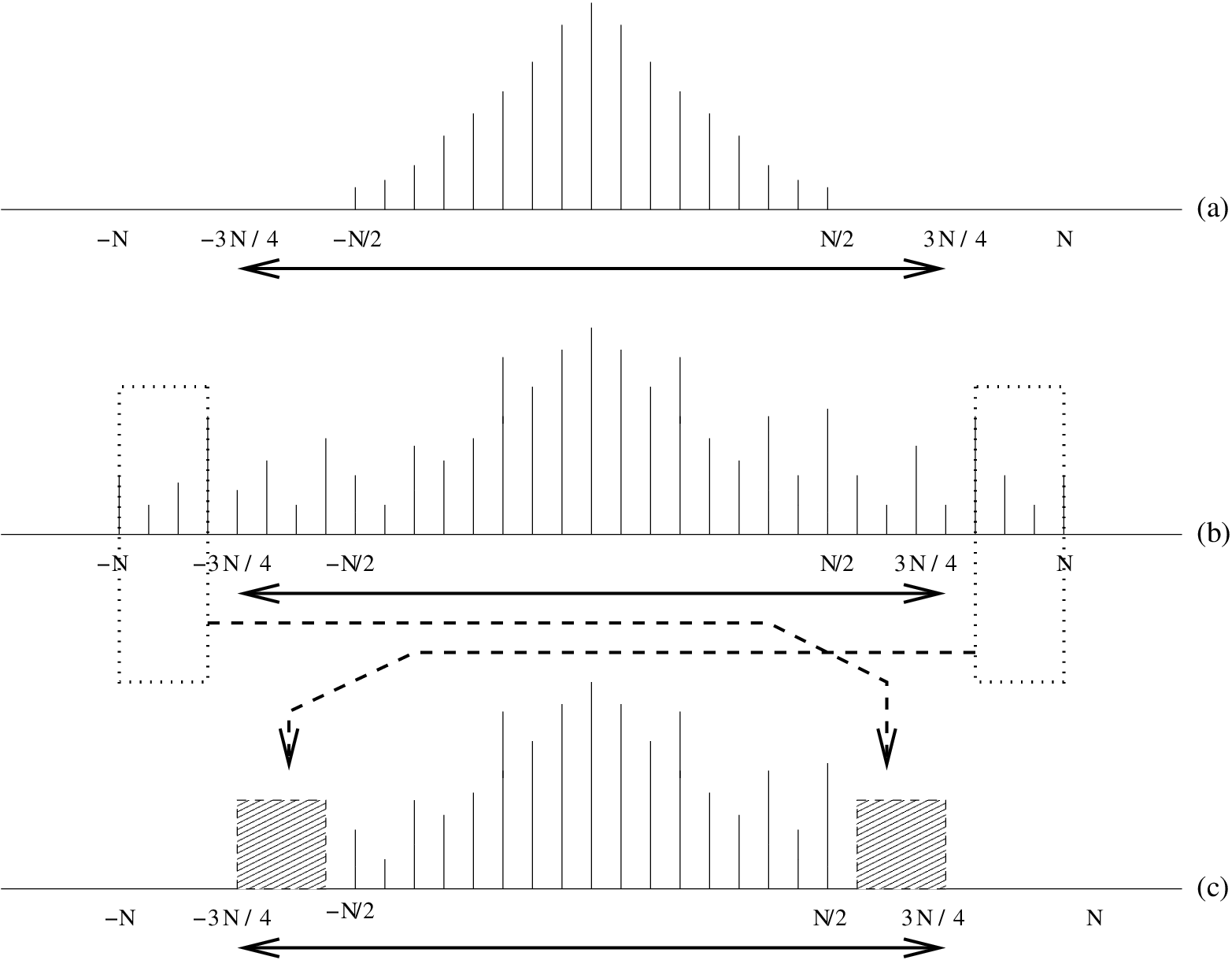}
\caption{Solving the aliasing problem with $M=3/2N+1$.  The arrow line
indicates the considered modes. (a) The original field $ h(x,t) $; (b)
the true function $ 1/2 \left[ \partial_x h(x,t)\right]^2$; (c) the
truncated function associated to $ 1/2 \left[ \partial_x
h(x,t)\right]^2$ when $M=3/2N+1$ }.
\label{fig13}
\end{figure}

In the specific case considered, function $ u(x,t)$ equals $ 1/2
\left[ \partial_x h(x,t)\right]^2$ and only contains a finite number
of non vanishing modes: to be precise, the $2N+1$ modes such that $ {
q}_{k} \equiv 2\pi k/L $ with $ k=-N, \cdots,N$ (see Fig.~\ref{fig12}
and \ref{fig13}). Observe that we are interested only in the modes
with $|k| \le N/2$. If we use for step (d),
\begin{itemize}
\item a truncated function with $M=N+1$ modes ($P=N/2$) to approximate
$ 1/2 \left[ \partial_x h(x,t)\right]^2$, all the Fourier components
$\widehat{h}_{ q_{k}} $ are modified.

\item a truncated function with $M=2N+1$ modes ($P=N$) to approximate
$ 1/2 \left[ \partial_x h(x,t)\right]^2$, all the non vanishing
Fourier components $\widehat{h}_{ q_{k}} $ ( $k=-N, \cdots,N$) are
correct, even those corresponding to spurious modes $ k=-N,
\cdots,-N/2-1$~~$ k =N/2 +1 , \cdots,N$.  By definition
$\widehat{h}_{{ q}_{k+rM}}=0$ for $r \ne 0$ hence the infinite sum in
(\ref{fourdimension5}) is reduced to the single term $r=0$.

\item a truncated function with $M=3N/2 +1 $ modes ($P=3N/4$) to
approximate $ 1/2 \left[ \partial_x h( x,t)\right]^2$, the Fourier
modes of the gradient square $ { q}_{p} = 2\pi p/L $ are correctly
obtained for $ k=-N/2, \cdots,N/2$ and spurious modes $ k=-N,
\cdots,-N/2-1$;~~ $ k=N/2 +1 , \cdots,N$ are not but this is of no
significance.  Again by definition $\widehat{h}_{{ q}_{k+rM}}=0$ for
$r \ne 0$ and $ |k| \le N/2 $ hence the infinite sum in
(\ref{fourdimension5}) is reduced to the single term $r=0$ for $ |k|
\le N/2 $.
\end{itemize}

\noindent As a consequence the choice $M=3N/2 +1$~($P=3N/4$) for step
(d) is the more economical.

\section{Correlation function and roughness}
\label{sec:appendixb}
By definition one has:
\begin{eqnarray}
G_2^2(r,t)&=& \frac{1}{V} \int_V~ d^{d}{\bf x} \left \langle
\left[ h({\bf x}+{\bf r},t)-h({\bf x},t) \right]^2 \right \rangle.
\label{appb_1}
\end{eqnarray}
Expanding this expression and integrating over ${\bf r}$ one gets
\begin{widetext}
\begin{eqnarray}
\frac{1}{V} \int_V~ d^{d}{\bf }r G_2^2(r,t) &=& W^2(t,L) +
\frac{1}{V^2} \int_V ~d^{d}{\bf x}~d^{d}{\bf r} \left \langle h({\bf
x},t)~h({\bf r},t) \right \rangle \\ \nonumber && + \frac{1}{V^2}
\int_V ~d^{d}{\bf x}~d^{d}{\bf r} \left \langle h^2({\bf x}+{\bf r},t)
\right \rangle -\frac{2}{V^2} \int_V ~d^{d}{\bf x}~d^{d}{\bf r} \left
\langle h({\bf x}+{\bf r},t)~h({\bf x},t) \right \rangle.
\label{appb_2}
\end{eqnarray}
\end{widetext}
For large $V$ we thus find
\begin{eqnarray}
W^2(t,L)&=& \frac{1}{2V} \int_V d^{d}{\bf r} G_2^2(r,t).
\label{appb_3}
\end{eqnarray}
It is then easy to show that if Eq.~(\ref{eq3.2}) and
Eq.~(\ref{eq3.3}) hold true, then the above relation implies that
Eq.~(\ref{eq3.5}) is valid.

\section{Scaling of the stationary structure factor}
\label{sec:appendixc}
Using the Fourier expansion, we get:
\begin{eqnarray}
h({\bf x}+{\bf r},t) &-&h({\bf x},t) =\nonumber\\&& \frac{1}{V} \sum_{{\bf k}}
\widehat{h}_{{\bf k}}(t) \left[ e^{i {\bf k}\cdot ({\bf x}+{\bf r})}-
e^{i {\bf k}\cdot {\bf x}} \right].
\label{appc_1}
\end{eqnarray}
By substituting in $G_2^2(r,t)$ one gets after few simple
manipulations
\begin{eqnarray}
&&G_2^2(r,t)=\nonumber\\&& \frac{1}{V^2} \sum_{{\bf k}} \left \langle
\widehat{h}_{{\bf k}}(t)~\widehat{h}_{-{\bf k}}(t) \right \rangle 2
\left[ 1 - \cos\left({\bf k}\cdot {\bf r}\right) \right].
\label{appc_2}
\end{eqnarray}
Note that, in the above equation, the term ${\bf k}=0$ yields a
vanishing contribution to the sum. Eq.~(\ref{appc_2}) can then be
inserted in Eq.~(\ref{appb_3}), yielding
\begin{eqnarray}
W^2(t,L)&=& \frac{1}{V^2} \sum_{{\bf k} \ne 0} \left \langle
\widehat{h}_{{\bf k}}(t)~\widehat{h}_{-{\bf k}}(t) \right \rangle.
\label{appc_3}
\end{eqnarray}
Therefore, since we define the structure factor from the relation
\begin{eqnarray}
\left \langle \widehat{h}_{{\bf k_{1}}}(t)~\widehat{h}_{{\bf
k_{2}}}(t) \right \rangle &=& V \delta_{{\bf k_{1}},-{\bf k_{2}}}
S(k,t),
\label{appc_4}
\end{eqnarray}
we find
\begin{eqnarray}
W^2(t,L)&=& \frac{1}{V} \sum_{{\bf k} \ne 0} S({\bf k},t).
\label{appc_5}
\end{eqnarray}
This relation can be checked to be valid directly from the definition
of $W^2(t,L)$. On the other hand from Eq.~(\ref{appc_2}) we also
obtain, using Eq.~(\ref{appc_4}),
\begin{eqnarray}
G_2^2(r,t) &=& \frac{2}{V} \sum_{{\bf k} \ne 0} S({\bf k},t) \left( 1
- e^{i {\bf k} \cdot {\bf r}} \right)
\label{appc_6}
\end{eqnarray}
or inverting
\begin{eqnarray}
\frac{1}{2} \int_V ~ d^{d} {\bf r} G_2^2(r,t) e^{-i {\bf k}\cdot {\bf
r}} &=& \nonumber\\\delta_{{\bf k},0} \left(\sum_{{\bf k^{\prime}}} S({\bf
k^{\prime}},t) \right)& - &S({\bf k},t).
\end{eqnarray}
Therefore for ${\bf k} \ne 0$ one gets
\begin{eqnarray}
 S({\bf k},t) &=& - \frac{1}{2} \int_{V} d^{d}{\bf r} G_2^2({\bf r},t)
e^{-i {\bf k}\cdot {\bf r}}.
\label{appb_7}
\end{eqnarray}
If we then assume that
\begin{eqnarray}
G_2^2(r,t)&=& r^{2\chi} g\left(\frac{t}{r^z}\right),
\label{appc_8}
\end{eqnarray}
and using the standard angular integral over the $d-$ dimensional
angular variables
\begin{eqnarray}
\int d \Omega_d~ e^{-i k r ~\cos\left(\hat{k}\cdot \hat{r}\right)} &=&
\left(2 \pi\right)^{d/2} \frac{J_{d/2-1} \left(k r
\right)}{\left(kr\right)^{d/2-1}},
\label{appc_9}
\end{eqnarray}
one easily gets
\begin{eqnarray}
S(k,t)&=& k^{-\left(d+2\chi\right)} \Phi\left(k^z~t\right),
\label{appc_10}
\end{eqnarray}
where
\begin{eqnarray}
&&\Phi\left(k^z~t\right)=\nonumber\\ &\hspace{-1cm}-&\hspace{-5mm}\frac{1}{2} \left(2 \pi\right)^{d/2}
\int_{0}^{kL} d \lambda \lambda^{2 \chi + d/2} g\left(
\frac{k^z~t}{\lambda^z}\right) J_{d/2-1} \left( \lambda \right).
\label{appc_11}
\end{eqnarray}
It is then immediate to see that
\begin{eqnarray}
\Phi\left(k^z~t\right)\sim \left\{ \matrix{ \mbox{const} \qquad
\mbox{for } k^zt \gg 1, \cr k^{d+2 \chi}~t^{2\beta+d/z} \qquad \mbox{for
} k^zt \ll 1. \cr} \right.
\label{appc_12}
\end{eqnarray}

\begin{table}[h]
\caption{Best estimates of critical exponents $\beta$, $\chi$ and $z$
in $d=2$ as reported in the literature. Note that in the numerical
solution of the continuum equation we have taken the value of the
non-linear coupling parameters corresponding more closely to the one
used in our work. Here KPZ stands for continuum KPZ equation with
finite-difference discretization, whereas FT stands for
field-theoretical methods. The error bars in the simulations are
typically of the order of the last reported digit.}

\begin{tabular}{lcccc}
\multicolumn{1}{l}{Model}&
\multicolumn{1}{c}{$\beta$}&
\multicolumn{1}{c}{$\chi$}&
\multicolumn{1}{c}{$z$}&
\multicolumn{1}{c}{Reference}\\
\hline
RSOS  & 0.245 & 0.393 & 1.607 & \cite{Marinari00} \\
RSOS  & 0.240 &   -   &   -   & \cite{Ala-Nissila94}\\
KPZ   & 0.25  & 0.39  &   -   & \cite{Amar90}\\
KPZ   & 0.240 &   -   &   -   & \cite{Moser91}\\
KPZ   & 0.240 & 0.404 &   -   & \cite{Beccaria94}\\
Flory & 0.25  & 0.4   &  1.6  & \cite{Kim89,Hentschel91}\\
FT    & 0.25  & 0.4   &  1.6  & \cite{Lassing98}\\
\hline
\end{tabular}
\label{table1}
\end{table}
\printtables
\newpage
\printfigures


\begin{thebibliography}{910}
\bibitem{note1} We note a misprint in Eq.~(8) of Ref. \cite{Giacometti01}:
 $\left( h_i-h_{i+1}\right)^2$ in the second term should be replaced
by $\left( h_i-h_{i-1}\right)^2$.
\bibitem{note2} 
As noted in~\cite{Mannella89}, a more sophisticated procedure using
two time steps is not necessarily more efficient for Langevin
equations. Indeed we tried some computations with the Heun
scheme \cite{Kloeden92}, an analogue of the Runge Kutta method for stochastic
equations~: it doubles the CPU time without increasing the accuracy.
\bibitem{note3} Since we are dealing with a continuum equation, any
quantity computed at the level of the lattice spacing, makes little
sense.
\end{thebibliography}
\end{document}